# Atomization modes for levitating emulsified droplets undergoing phase change

D. Chaitanya Kumar Rao[*] and Saptarshi Basu[§]

*Department of Mechanical Engineering, Indian Institute of Science, Bangalore – 560012, India*

## Abstract

We delineate and examine the distinct breakup modes of evaporating water-in-oil emulsion droplets under acoustic levitation. The emulsion droplets consist of decane/dodecane/tetradecane as oil, while the water concentration is varied from 10% to 30% (v/v). The droplets were heated under different laser irradiation intensities and were observed to exhibit three broad breakup mechanisms, viz., breakup through bubble growth, sheet breakup, and catastrophic breakup. The occurrence of these modes of the breakup is found to be primarily dependent on the volatility differential among the components. Early nucleation in water/decane emulsions results in the growth of vapor bubble, which is characterized by intricate patterns of wave propagation on the droplet surface. The formation of these patterns suggests that the short time scale and length scale of wave patterns is the manifestation of Faraday instability, triggered on the droplet surface by the acoustic field induced resonance. A sheet-like breakup, on the contrary, occurs predominantly in emulsions comprising of components with relatively high volatility difference (water/dodecane emulsions) due to the breakup of an indiscernible small sized bubble. Intense catastrophic breakup occurs for emulsions with significantly vast volatility difference (water/tetradecane emulsions) where the droplet undergoes prompt fragmentation into fine secondary droplets.

## 1 Introduction

The atomization of liquids is one of the most fundamental fluidic phenomena, which has been a subject of extensive scientific research and technological applications (such as ink-jet printing, combustion engines, and irrigation) (Eggers and Villermaux 2008). For example, in automotive sprays, the creation of large droplets is usually undesirable. These large droplets vaporize slowly and result in a locally rich/lean mixtures, which in turn lead to thermoacoustic instabilities, combustion inefficiency, and unwanted pollutant emissions. Recently, emulsified mixtures, which are multi-component multiphase systems, have stimulated considerable interest owing to their practical significance and cross-cutting influence across disciplines. In particular, emulsions are known for their potential benefits in gas turbine combustors and internal combustion engines for improving the combustion efficiency along with reducing the pollutant emissions (e.g., NOx and particulate matter) during the combustion process (Kadota and Yamasaki 2002). In a typical water-in-oil emulsion, water exists in the form of dispersed small sub-droplets. In addition, since the volatile dispersed phase (water) and the continuous phase (oil) form an immiscible mixture, the two phases vaporize independently. Therefore, during the evaporation or combustion of an emulsified droplet, the temperature of the droplet can reach the superheat limit of the water (since the boiling temperature of the oil is higher than that of water). Due to the superheated metastable state of water, embryonic nuclei are generated in several sites, which may consequently coalesce and grow with time. The breakup of these bubbles (puffing/micro-explosion) leads to the fragmentation of parent droplet into multiple secondary droplets. This explosive boiling of the dispersed water sub-droplets and the resulting breakup of the whole droplet is referred to as puffing/micro-explosion or secondary atomization (Avedisian and Andres 1978; Law et al. 1980; Mikami et al. 1998; Chen and Lin 2011; Watanabe and Okazaki 2013).

[*]chaithanyadevv@gmail.com

[§]sbasu@iisc.ac.in

In both experiments as well as theoretical modeling, it has been verified that the evaporation/combustion behavior of emulsified droplets is primarily dependent on the relative volatilities and volume fractions of water and fuel components, internal circulation, and the immiscibility of water and fuel (Law et al. 1980; Lasheras et al. 1981b, 1984). Moreover, it has been shown that an increase in water concentration, as well as fuel boiling temperature, increases the probability and strength of droplet breakup (Lasheras et al. 1981b; Law and Law 1982). Jackson and Avedisian (1998) also reported that an increase in water concentration lowers the probability of soot formation during the combustion of water-in-heptane emulsions. Recently, Shinjo et al. (2014) modeled the physics underlying the breakup of water-in-oil emulsion droplet via interface-capturing. It was shown that the size and location of the water sub-droplet determine the bubble growth and breakup dynamics. In particular, it was reported that when the size of sub-droplet is small, the breakup of parent droplet is limited. Moreover, it was shown that the interaction between multiple breakup events increases the degree of breakup. The breakup of emulsified droplets has been studied using both intrusive techniques (suspending fiber or thermocouple) (Kimura et al. 1986; Watanabe et al. 2010; Califano et al. 2014; Khan et al. 2014; Mura et al. 2014; Kim and Baek 2016) and low or non-intrusive methods (such as freely falling approach under microgravity conditions) (Lasheras et al. 1979, 1981a; Jackson and Avedisian 1998; Matsumoto et al. 1999; Segawa et al. 2000). In the case of intrusive techniques, the experiment is influenced by the interaction of droplet with the fiber or thermocouple, which in turn can lead to the creation of undesirable heterogeneous nucleation sites within the droplet (Law 2010).

On the other hand, the freely falling technique has been useful in the accurate determination of parameters such as droplet burning-rate, flame diameter, and soot behavior (Shaw et al. 1988; Dietrich et al. 1996). However, the complexity involved in the freely falling droplet technique does not allow one to qualitatively evaluate the complex processes such as the instabilities at the vapor-liquid interface. Similar to freely falling droplet technique, the low intrusive or contact-free nature of acoustic levitation (Shitanishi et al. 2014; Gonzalez Avila and Ohl 2016; Hasegawa et al. 2016; Pathak and Basu 2016a; Pathak et al. 2017; Zang et al. 2017; Di et al. 2018; Kobayashi et al. 2018; Raju et al. 2018; Watanabe et al. 2018) along with its uncomplicated experimental apparatus allows one to capture short spatio-temporal instabilities such as Kelvin-Helmholtz or Rayleigh-Taylor instabilities with reduced difficulty (Basu et al. 2012; Basu et al. 2013; Gonzalez Avila and Ohl 2016).

Although the existing experimental investigations (primarily in combustion) are comprehensive in terms of understanding the influence of emulsion type (e.g., water-in-oil or oil-in-water), size of water sub-droplet and coalescence on the breakup of droplets, the literature is mostly devoid of any study concerning the delineation and comprehensive analysis of the different types of droplet breakup mechanisms, especially during evaporation. In-depth studies on the atomization of emulsions are rare, especially under contact-free conditions like acoustic levitation. Moreover, although the occurrence of breakup is known in emulsions, no previous studies have delineated different modes and regimes in breakup or reported phenomena such as the propagation of capillary waves and Faraday instability or performed a comprehensive experimental and theoretical analysis of ligament-mediated atomization. This study is an effort to address questions such as what role does the size of the vapor bubble plays in determining the mode of breakup? Alternatively, how does the ligament-mediated breakup differ in emulsified droplets compared to miscible multi-component droplets?

In the present work, the breakup of the levitated emulsified droplets is achieved by externally heating the droplet at different laser irradiation intensities. The breakup modes are broadly categorized into three types, namely, bubble breakup, sheet breakup, and catastrophic breakup. Section 3.1 provides an overview of breakup mechanisms observed in the water-in-fuel emulsified droplets. Section 3.2 lays out different regimes during the bubble growth and ligament-mediated breakup of droplets along with theoretical validation. The first observation of wave propagation and Faraday instability during vapor bubble growth in emulsified droplets is reported. Section 3.3 details the different categories of sheet formation and breakup. Finally, section 3.4 describes the occurrence of catastrophic breakup due to a significantly high volatility difference among the fuel components. The influence of volatility differential on the onset and mode of the breakup is described. A regime map is proposed, which shows a relation between the ratio of force generated by bubble and surface tension to the size of the secondary droplets created from droplet breakup.

## 2 Experimental methodology

A single micron-sized emulsion droplet (400 ± 25 µm) is levitated at a pressure node of the standing-wave generated by a single-axis acoustic levitator (Model: Tec5, 100 kHz frequency, 154 dB). The levitated droplets are heated using a continuous wave $CO_2$ laser (Model: Synrad 48, wavelength of 10.6 μm, and beam diameter of 3.5 mm) at five different laser intensities, viz., 3.5, 5, 6.5, 8, and 9.5 W. No lenses were used to focus the laser beam. In all the test cases, the laser remained turned on for at least 5 s, which is a significantly longer period compared to the total evaporation and breakup time of the droplet (~ 0.5 s).

The emulsion mixtures consist of distilled water (dispersed phase), while three different hydrocarbons viz., decane, dodecane, and tetradecane are chosen as the heavier components (continuous phase). Total nine mixtures of water/decane, water/dodecane and water/tetradecane are considered, viz.: (a) 10% by volume water in decane/dodecane/tetradecane (DW10/DDW10/TDW10), (b) 20% by volume water in decane/dodecane/tetradecane (DW20/DDW20/TDW20), and (c) 30% by volume water in decane/dodecane/tetradecane (DW30/DDW30/TDW30). A constant 2.5% (v/v) surfactant (SPAN 80) is used to stabilize the emulsion. SPAN 80 (Sorbitan Monooleate) was chosen as an emulsifying agent since its HLB (hydrophile-lipophile balance) is close to the hydrocarbons used in the present work (HLB – 4.3), and it has been identified that the emulsification is efficient when the surfactant and the hydrocarbons have similar HLB (Chen and Tao 2005). For preparing the emulsions, the hydrocarbon fuel and the surfactant are first mixed, and then water is added gradually until emulsions of milky white appearance are formed. These emulsified mixtures are sonicated for 15 minutes with continuous stirring and are found to be stable for at least half an hour at room temperature (30 °C). A similar procedure for emulsion preparation was used by Kim and Baek (2016). It is well acknowledged that emulsions exhibit non-Newtonian behaviors when the concentration of the dispersed phase is significant. Therefore, to avoid the influence of non-Newtonian behavior on the breakup phenomena, only dilute emulsions (up to 30% (v/v) water) are considered in this study. It is important to note that the coalescence of water sub-droplets may occur during the heating of the emulsion droplet (Kadota et al. 2007; Mura et al. 2014). However, it is rather difficult to quantitatively compare the process of coalescence since these studies involve the combustion of significantly larger droplets (> 1 mm) at temperatures of the order of 800 °C. In contrast, the present study involves the evaporation of relatively small droplets under laser irradiation. Nevertheless, the influence of solubility or coalescence on the breakup modes is not elaborated in the present study, which is beyond the scope of present work.

The evaporating back-illuminated droplets are visualized using a CMOS high-speed camera (Phantom Miro Lab110) coupled with a 5X zoom lens. The schematic of the experimental setup illustrating the isometric view (Fig. 1(a)) and top view (Fig. 1(b)) is shown in Fig. 1. The images are obtained at spatial resolutions of 6.6 µm/pixel (high magnification) and 19 µm/pixel (low magnification) at a frame rate of 20 kHz with the exposure maintained at 10 µs. An Infrared camera was used to measure the droplet temperature. The IR camera (FLIR SC5200: pre-calibrated for a standard emissivity of 1 with an accuracy of ± 1 °C) was operated at 500 frames per second (fps) with a spatial resolution of 12.4 µm/pixel. It has been reported that the emissivity for water is between 0.95 and 0.98 (Bremson 1968; Wolfe and Zissis 1985). The change in temperature due to the change in emissivity is 0.03 °C, which is assumed to be negligible.

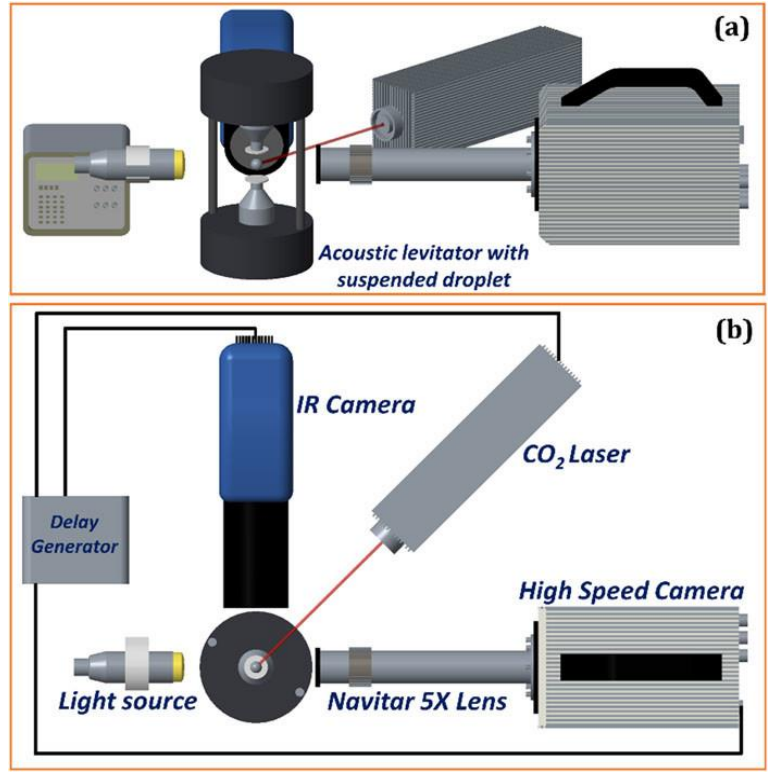

FIG. 1. Schematic showing (a) isometric view and (b) top view of the experimental apparatus, adapted from Pandey and Basu (2018).

The captured images were processed using ALTAIR software (FLIR Systems, version 5.91.010) to extract the droplet temperature during the heating process. The temperature information was gained by defining a linear region of interest along the droplet diameter in each IR frame, and the average temperature on the surface of the droplet was calculated. The temperature at the core of the droplet is anticipated to be higher than that at the surface due to penetration of radiation inside the droplet (Abramzon and Sazhin 2006; Pathak and Basu 2016a). Due to the relatively small depth of field, the IR images were sometimes out of focus. Therefore, only the images which were in focus were considered for further analysis. The uncertainty in the measurement of droplet surface temperature is ± 1 °C. The synchronization of the high-speed camera and Infrared camera with the laser is achieved using a delay generator (BNC 575). The diameter of the water sub-droplets in the emulsion is measured using an optical microscope (Olympus BX-51, Japan). The observation of emulsions under the microscope showed that the diameter of water sub-droplets is rich in the range of 10–50 µm (~65%), followed by droplets in the range of 50–90 µm (~15%) and 90–130 µm (~15%). The sub-droplet sizes >130 µm are observed to be least probable (~5%) (see Fig. 2).

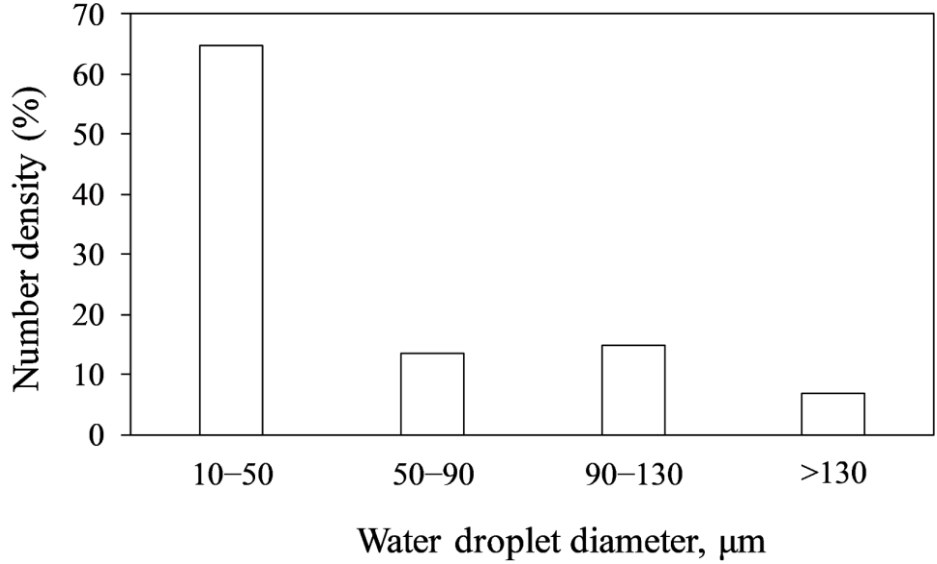

FIG. 2: Size distribution of water sub-droplets inside a typical emulsion mixture (30% (v/v) water-in-decane).

The evolution of droplet diameters is obtained from frame-by-frame analysis with an in-house MATLAB code (Rao and Karmakar 2018). Similarly, an image analysis program, Image-Pro Plus (Media Cybernetics, Inc.) is employed to measure the velocity and acceleration of residual and secondary droplets, bubble diameter, ligament diameter, and mean diameter of the secondary droplets (which is the average length of diameters measured at 2-degree intervals and passing through droplet's centroid). The experimental errors (primarily uncertainty in pixel identification), together with calibration uncertainties, are accounted for no more than 2% of the bubble, ligament, and secondary droplet diameter determination. It is important to note that the breakup phenomena of emulsions exhibit stochastic behavior primarily due to two uncontrollable reasons (i) spatial distribution of the dispersed phase during emulsion preparation and (ii) randomness of nucleation sites. Besides, proper care has been taken to minimize the asymmetry in the initial drop placement. Therefore, only the most probable cases were considered for further analysis. The experiments were repeated at least 20 times for each emulsion to ensure the repeatability of the reported phenomena.

# 3 Results and discussion

## 3.1 Overview

In water-in-fuel emulsion, water exists as stationary sub-droplets inside the continuous phase (fuel). Therefore, in an emulsion droplet, the volatile constituent (water) can be heated beyond its boiling temperature (referred to as superheating) since evaporation from the droplet surface is principally achieved by the low volatile continuous phase (fuel). In essence, the higher volatile component (water) undergoes phase transformation, while the lower volatile component (fuel) plays a role in heating up the former. Since relatively high boiling point fuels are chosen in this study (decane, dodecane, and tetradecane), the dispersed water sub-droplets are superheated. This metastable state is preserved until the rapid vaporization of water sub-droplets and subsequent homogeneous nucleation of vapor bubbles inside the droplet. Consequently, the vapor bubble undergoes fragmentation with or without its growth inside the parent droplet.

In particular, for water/decane emulsified droplets, since the volatility differential between decane and water is small, the droplet breakup strength is limited. However, due to the relatively higher volatility of decane (compared to dodecane and tetradecane) and since lower volatile component in an emulsion controls the temperature of the dispersed phase, the water component is superheated sufficiently early in droplet's lifetime. It can be intuitively argued that greater the volatility disparity between the two liquids, higher is the breakup strength.

In contrast, the high-intensity breakup for tetradecane-based emulsion droplets completely disintegrates the parent droplet into a plethora of fine fragments. The breakup phenomenon in these tetradecane emulsions is anticipated to be triggered due to the large volatility differential between tetradecane and water (compared to decane and dodecane emulsions), resulting in a higher degree of superheat. Unlike decane emulsions, the onset of the breakup is delayed in tetradecane emulsions where nucleation and subsequent droplet fragmentation occurs near the end of its lifetime. Despite the delayed fragmentation, the overall droplet lifetime is substantially reduced due to significant strength of the breakup phenomenon resulting in small-sized secondary droplets (~ 20 µm). It is important to note that, in the present study, the increase in the concentration of water was not observed to play an essential role in the breakup phenomenon except for a marginal shortening of the droplet lifetime. This is possibly due to the reason that there is no discernible change in the size of sub-droplets for different concentrations of water in the emulsion. A similar observation was reported by Kim and Baek (2016) for different water concentrations in water-in-decane emulsions.

Figure 3(a)−(c) shows the evolution of droplet surface temperature for DW30, DDW30, and TDW30, respectively, for five different laser powers. The temperature profile for all the droplets consists of two predominant stages, 1) the initial transient regime (preheating in Fig. 3) through which the bulk temperature increases and 2) the steady-state regime (steady regression in Fig. 3) through which the temperature is constant (wet bulb temperature). The time and temperature scales associated with the external heating of a droplet is determined by both the phases. In the preheating regime, the history of droplet temperature is dictated by a balance between the energy provided by the heating source (laser) and the sensible enthalpy required for increase in the temperature of the droplet. In the steady-state regime, the droplet exhibits a steady-state temperature (wet bulb limit) with all the input energy from the external heating source being utilized for vaporization (steady regression regime). The breakup phenomena shown in the present work occur only during the steady-state regime. Thus, it is plausible that coalescence may occur during the preheating regime. However, the influence of solubility or coalescence on the breakup modes is not elaborated in the present work, which needs a separate in-depth investigation.

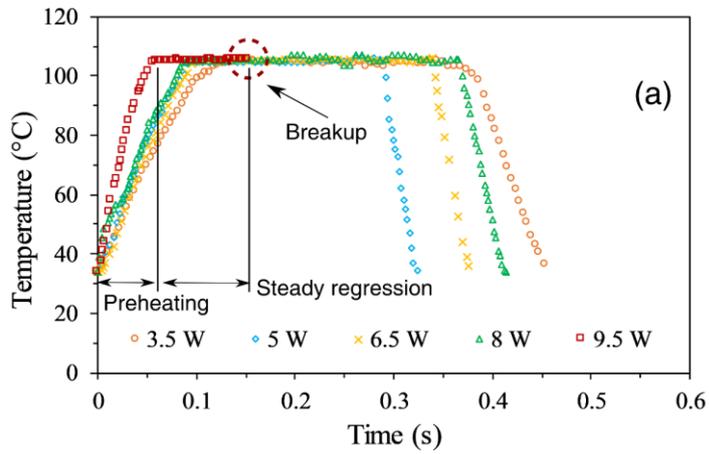

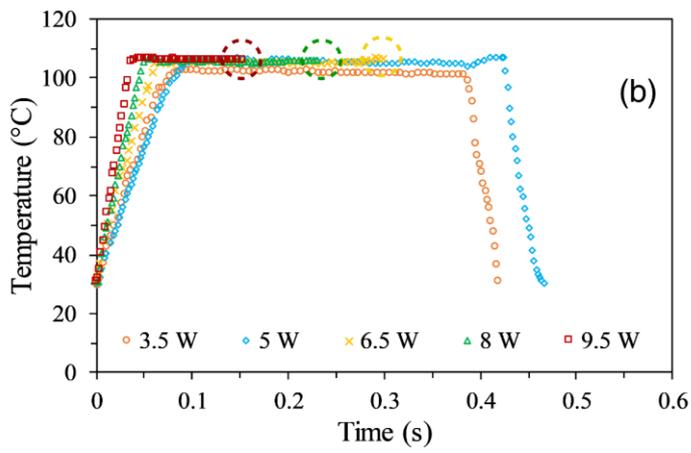

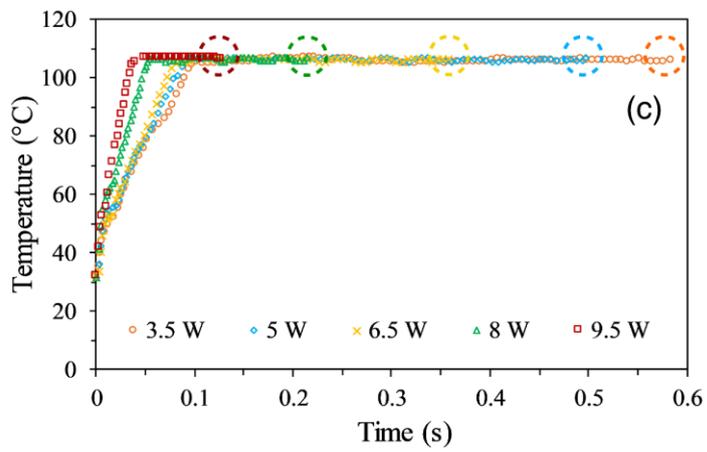

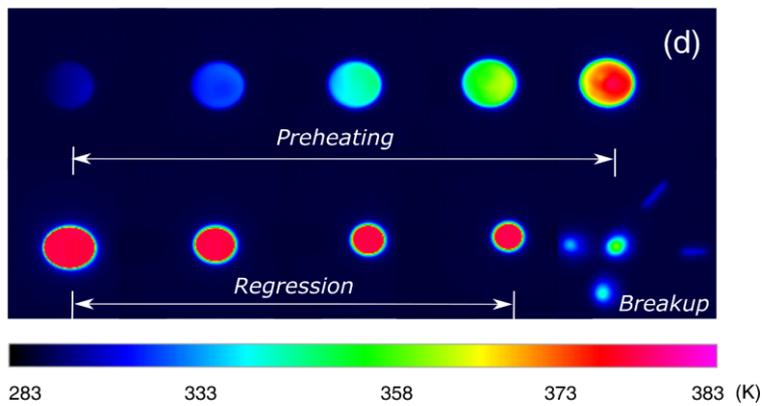

FIG. 3. Temporal evolution of the droplet surface temperature corresponding to (a) DW30, (b) DDW30, and (c) TDW30 droplets at different laser heating intensities. The dotted circles identify the onset of droplet breakup. Each plot represents the average of at least three test runs with a maximum error of ± 2 °C. (d) Infrared images showing the droplet surface temperature distribution during the preheating and regression regimes followed by the breakup of a typical DDW30 droplet at 9.5 W laser power.

As seen in Fig. 3, with the increase in laser heating intensity, the rate of preheating increases, but the droplet surface temperature during the steady regression regime remains constant for all the emulsions and laser intensities. A higher heating rate plays a significant role in the nucleation of the vapor bubble and consequent breakup of the droplet at the end of the steady regression regime. However, whether the droplet undergoes fragmentation or not depends on the laser irradiation intensity as well as the constituents present in the emulsion. The dotted circles in Fig. 3(a)−(c) indicate breakup of droplet for respective laser intensities. It is evident that, unlike decane and dodecane emulsions, tetradecane emulsion droplet undergoes breakup at all the laser irradiation intensities. Fig. 3(d) shows the droplet surface temperature distribution of a typical emulsion droplet (DDW30) at high laser power (9.5 W), which is characterized by preheating, regression, and breakup of the droplet.

It is important to note that, for the emulsion droplet with non-breakup cases (low heating powers), the occurrence of any instability or disruptive phenomena was not observed; instead, the droplet smoothly evaporates with time. However, in a few cases (although with an extremely low probability of < 5% test cases) flattening of the parent droplet occurs at the end of droplet lifetime which then leads to the sudden fragmentation of the droplet due to the domination of the acoustic pressure over the surface tension of the liquid (Pathak and Basu 2016a).

Table 1 shows the dominant modes of breakup for various laser powers and emulsified mixtures studied in this work. Among the different laser powers used to heat the emulsified droplets, heating at 9.5 W results in the maximum probability of breakup for all the droplets. Apart from the differences in the probability, no discernible variation in the strength of breakup was witnessed for different laser intensities. This implies that once the breakup criterion is satisfied, the breakup intensity is largely decoupled from the heating rate. Therefore, further analysis of the breakup phenomena will be focused on the droplets heated at 9.5 W.

Table 1. Essential characteristics of breakup phenomena in different emulsified mixtures.

| Mixtures | Dominant breakup mode | Probability of breakup (%) | | | | |
|---|---|---|---|---|---|---|
| | | 3.5 W | 5 W | 6.5 W | 8 W | 9.5 W |
| DW10 | Bubble growth or Puffing | 0 | 5 | 25 | 70 | 100 |
| DW20 | Bubble growth/sheet breakup | 0 | 15 | 40 | 65 | 100 |
| DW30 | Bubble growth/sheet breakup | 0 | 30 | 40 | 70 | 100 |
| DDW10 | Sheet breakup/bubble growth | 0 | 15 | 35 | 60 | 100 |
| DDW20 | Sheet breakup/bubble growth | 0 | 15 | 45 | 70 | 100 |
| DDW30 | Sheet breakup/bubble growth | 0 | 35 | 60 | 75 | 100 |
| TDW10 | Catastrophic breakup | 40 | 30 | 45 | 85 | 100 |
| TDW20 | Catastrophic breakup | 55 | 55 | 65 | 100 | 100 |
| TDW30 | Catastrophic breakup | 60 | 70 | 85 | 100 | 100 |

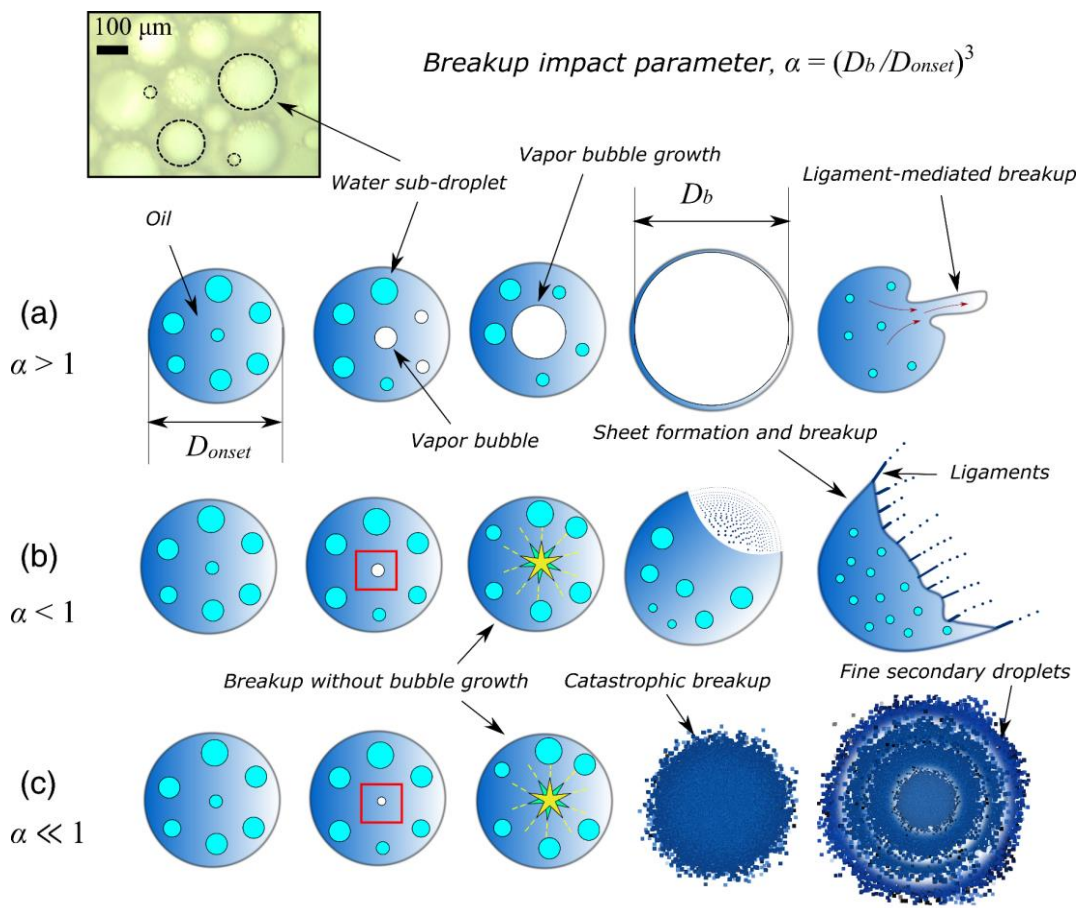

FIG. 4. Schematic showing the notional pathways of (a) Breakup through bubble growth, (b) sheet breakup, and (c) catastrophic breakup. The inset on the left side of the figure shows water sub-droplets under a microscope.

Figure 4 illustrates the notional pathways of the different breakup mechanisms observed in the present study. The bubble breakup resulting in different breakup mechanisms is quantified using a breakup impact parameter ($\alpha$), which is expressed as (Rao et al. 2018)

$$\alpha(t) = (\frac{\pi}{6} D_b^3) \Big/ (\frac{\pi}{6} D_{onset}^3), \qquad (1)$$

Here $D_b$ and $D_{onset}$ represent the bubble diameter at the pre-breakup instant and droplet diameter at the onset of nucleation, respectively. Unlike the miscible droplets (with significant volatility differential among the components) (Rao et al. 2017a, 2018), emulsified droplets exhibit both larger $D_b$ (i.e., $\alpha > 1$) as well as significantly smaller $D_b$ (usually indiscernible) compared to $D_{onset}$ (i.e., $\alpha \ll 1$). Due to the relatively lower volatility differential between decane and water, the size of the bubble at the onset of breakup ($D_b$) is large (~ 1 mm) (illustrated in Fig. 4(a)). However, a more substantial difference in the volatility between tetradecane and water causes the breakup of an indiscernible small sized bubble (< 0.01 mm) soon after its formation (Fig. 4(c)).

A large $\alpha$ usually implies that the net force acting at the vapor-liquid interface is always more significant than the capillary restoring force of the liquid film surrounding the bubble. This restoring force subsequently fails to withstand the net pressure due to bubble expansion. Consequently, the liquid layer surrounding the bubble turns thinner and ultimately ruptures into a stretching ligament and secondary droplets (Fig. 4(a)). In contrast, a significantly smaller $\alpha$ implies that the high-pressure vapor bubble cannot expand further inside the immiscible emulsion droplet, leading to its intense fragmentation via sheet or catastrophic breakup modes (Fig. 4(b)−(c)).

Although the occurrence of droplet breakup via homogeneous nucleation is a stochastic phenomenon (Mikami et al. 1998; Mikami and Kojima 2002), the prediction of breakup modes in emulsions is even more complex since the fuel and water components do not mix. Nevertheless, distinct breakup processes have been delineated for different water-in-fuel mixtures based on the prominent qualitative features and probability of bubble nucleation. The onset of bubble nucleation is determined using a parameter referred as normalized squared onset diameter (NOD), which is described as the square of the ratio of droplet diameter at the onset of nucleation ($D_{onset}$) to the initial droplet diameter ($D_0$) (Zeng and Lee 2007; Rao et al. 2017b). Figure 5 shows the experimental observation of three dominant modes of fragmentation during the evaporation of emulsified droplets, viz. bubble growth and breakup (Fig. 5(a)−(b)), sheet breakup (Fig. 5(c)−(d)), and catastrophic breakup (Fig. 5(e)). Figure 5(a) represents the dominant mode of bubble breakup via Faraday instability, which occurs when the bubble nucleates early in the lifetime of the droplet (NOD ~ 1). The breakup of droplets also occurs without the instability being triggered; however, this happens only when the nucleation is delayed (NOD ~ 0.2). Similarly, sheet breakup can be further categorized into stable sheet breakup and unstable sheet breakup depending on the onset of nucleation. Stable sheet breakup (Fig. 5(c)) is the dominant mode of sheet fragmentation (NOD ~ 0.8). Therefore, further discussion on unstable sheet fragmentation (Fig. 5(d)) will not be included in this paper. However, readers are referred to Rao et al. (2019) for a detailed study on unstable sheet fragmentation.

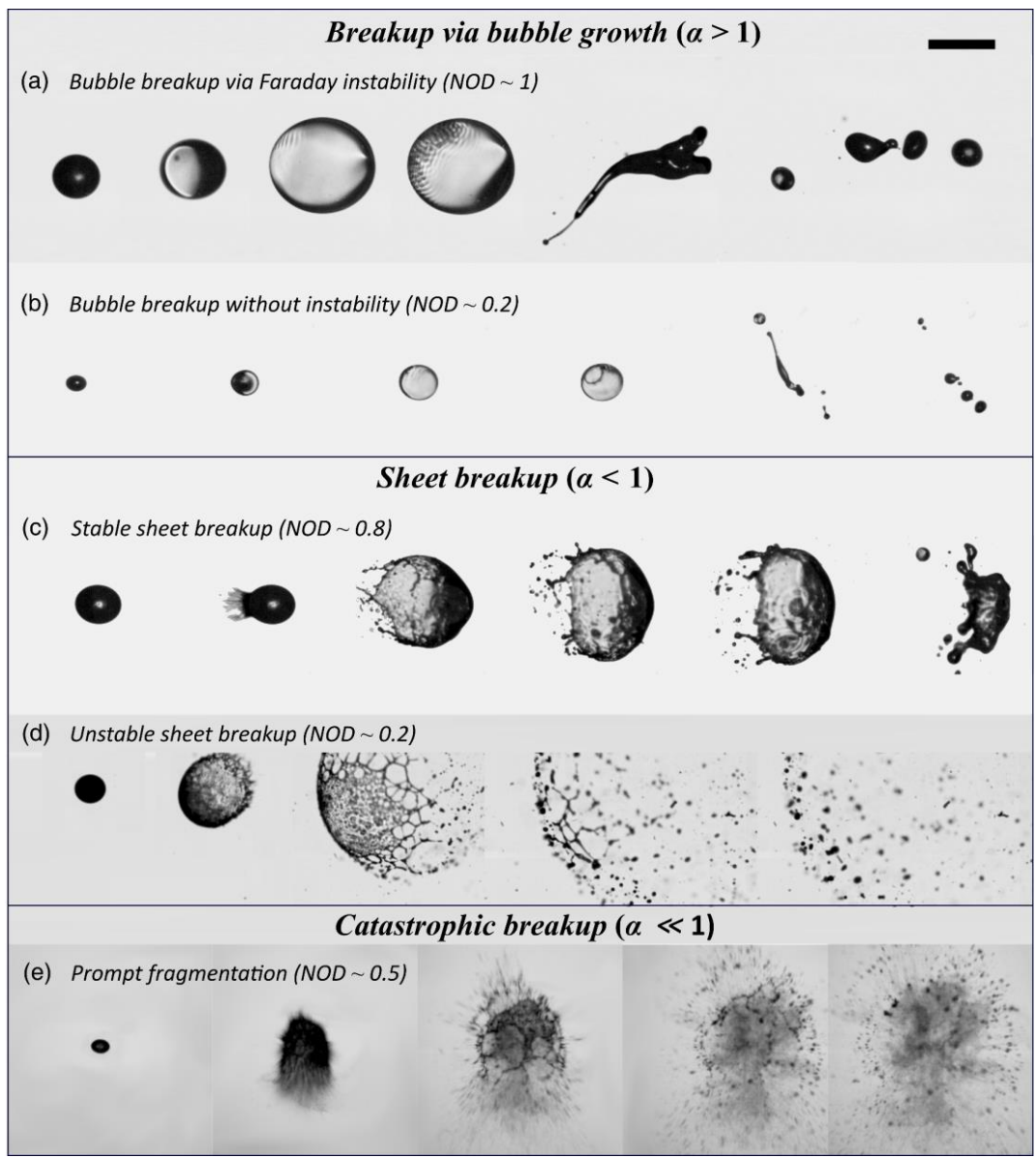

FIG. 5. Modes of fragmentation observed during the evaporation of water-in-fuel emulsion droplets. The scale bar represents 500 µm and it is identical across all the sequence of images.

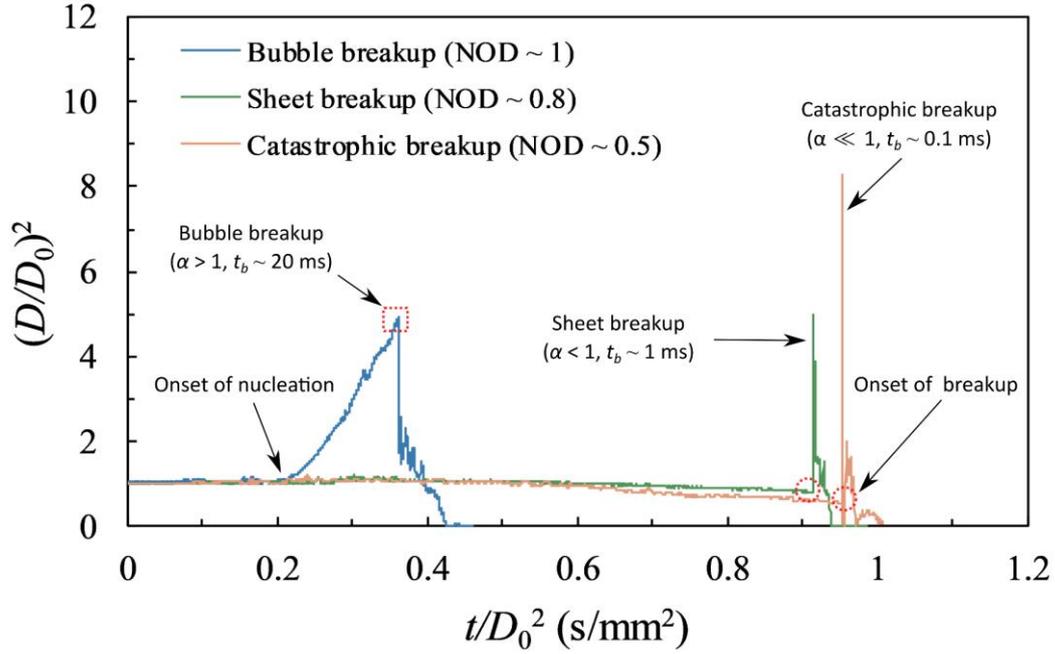

FIG. 6. Temporal evolution of droplet diameters resulting in bubble breakup, sheet breakup, and catastrophic breakup. $D_0$ represents initial droplet diameter.

The above-mentioned three prominent breakup modes can also be distinguished from the diameter regression profiles of the emulsion droplets. As can be seen in Fig. 6, the bubble breakup event ($\alpha > 1$) is associated with an early stage bubble nucleation (NOD ~ 1), which is followed by a relatively slow bubble growth ($t_b$ ~ 20 ms) until the bubble ruptures. The peak associated with the breakup event (highlighted by a red dotted rectangle) signifies the maximum bubble diameter before rupture. In contrast, delayed nucleation (and hence a high degree of superheat), is associated with sheet breakup (NOD ~ 0.8, $\alpha < 1$) and catastrophic breakup (NOD ~ 0.5, $\alpha \ll 1$), where the bubble undergoes instant fragmentation with no scope for bubble growth.

### 3.2 Breakup through bubble growth

#### 3.2.1 Nucleation and bubble growth

Among the several test fluids investigated in this study, the breakup of the droplet through the collapse of a large bubble ($\alpha > 1$) occurs predominantly in decane emulsions followed by dodecane mixtures. Due to volatility differential, water ($T_{b,\,water} = 373$ K) is heated beyond its boiling temperature by the transfer of heat from the surrounding fuel ($T_{b,\,decane} = 447$ K), resulting in the nucleation of bubbles within the droplet. Figure 7 shows the typical sequence of events demarcating bubble growth inside the droplet. The time listed beneath each frame is the elapsed time since the observation of the vapor bubble. Here $\Delta t = 0$ ms represents the onset of homogeneous bubble nucleation, assuming a negligible difference between the actual onset of nucleation and the first instance of vapor bubble observation. The vapor nuclei at several sites ultimately coalesce and develop into a single growing bubble (visible at $\Delta t = 4.8$ ms). The growth of a single bubble also indicates the absence of any undesirable contaminants leading to heterogeneous nucleation sites. The growth of the vapor bubble uniformly stretches the liquid. This stretching finally results in the bubble diameter exceeding $D_{onset}$ and reaches a maximum equivalent diameter of approximately 800 µm. The eccentricities in the position of the bubble become negligible with the growth of the vapor pocket. This is due to the fact that significant differences in the densities of the vapor and liquid result in rapid bubble growth with negligible liquid evaporation. Therefore, the fuel component (decane) develops into a thin layer surrounding the bubble (18.4 ms). During the majority of the bubble growth period, the surface of the droplet remains featureless. However, ripples begin to appear on the droplet surface (22.5 ms) in the final stages of bubble growth.

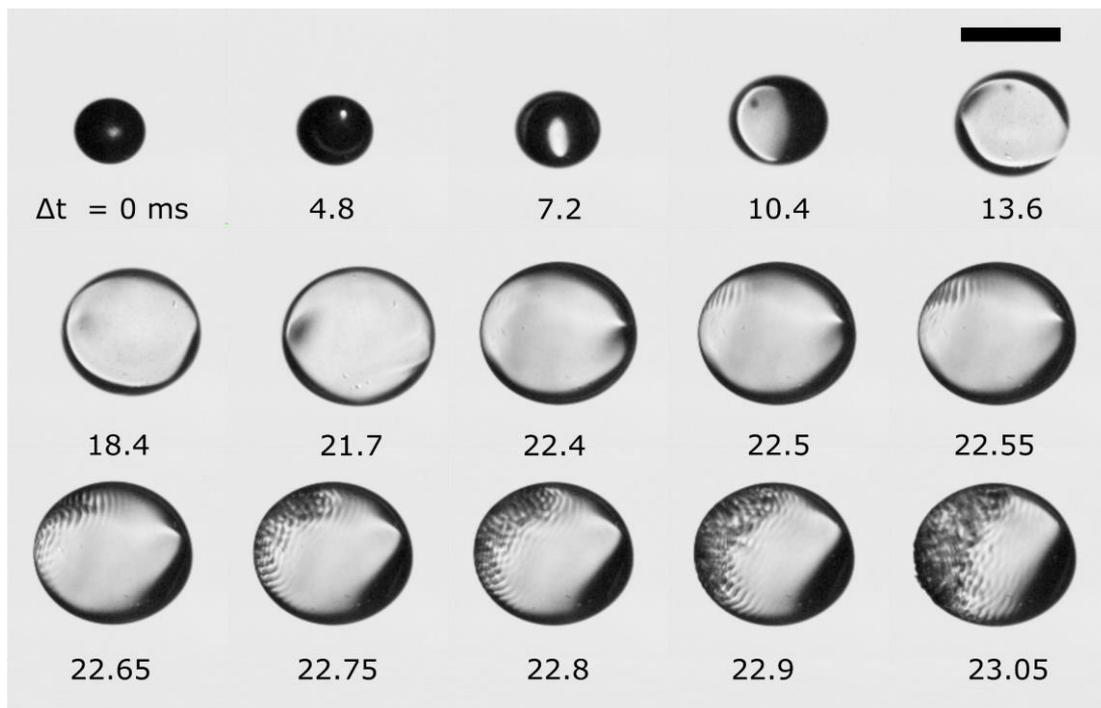

FIG. 7. The growth of the vapor bubble inside an emulsion droplet followed by the wave propagation on the surface of the droplet. The scale bar indicates 500 µm.

The influence of gravity on the droplet surface can be neglected during bubble growth since the wavelengths are significantly smaller ($45 \pm 5$ µm) than the bubble size at the onset of wave propagation (~ 800 µm). Therefore, it is highly probable that these waves are, in fact, capillary waves, initiated by the combined influence of bubble inertia and liquid surface tension. The propagation of similar capillary waves during bubble growth has also been reported previously in burning pendant multi-component droplets (Rao et al. 2018). However, unlike the previous study, there are no intrusive effects during the lifetime of the evaporating droplet.

Soon after the appearance of capillary ripples on the droplet surface, the waves grow in amplitude and complexity due to the possible influence of the acoustic field (22.8 ms). The formation of these complex wave structures on the droplet surface is anticipated to be the consequence of an interaction between the zonal capillary waves on the liquid film surrounding the vapor bubble and the meridional waves produced by the vibrational dynamics of the transducer perpendicular to the zonal waves (illustrated in Fig. 8(a)). The resulting oscillatory mode shapes on the droplet surface due to this vertical vibration can be attributed to the Faraday wave instability (Muller et al. 1998). These meridional waves, coupled with the zonal counterpart, lead to the creation of circular standing waves or craters on the droplet surface. The craters are initially stationary; however, due to the acoustic field, they gradually rotate along the periphery of the droplet in either clockwise or anti-clockwise direction until the breakup of the bubble.

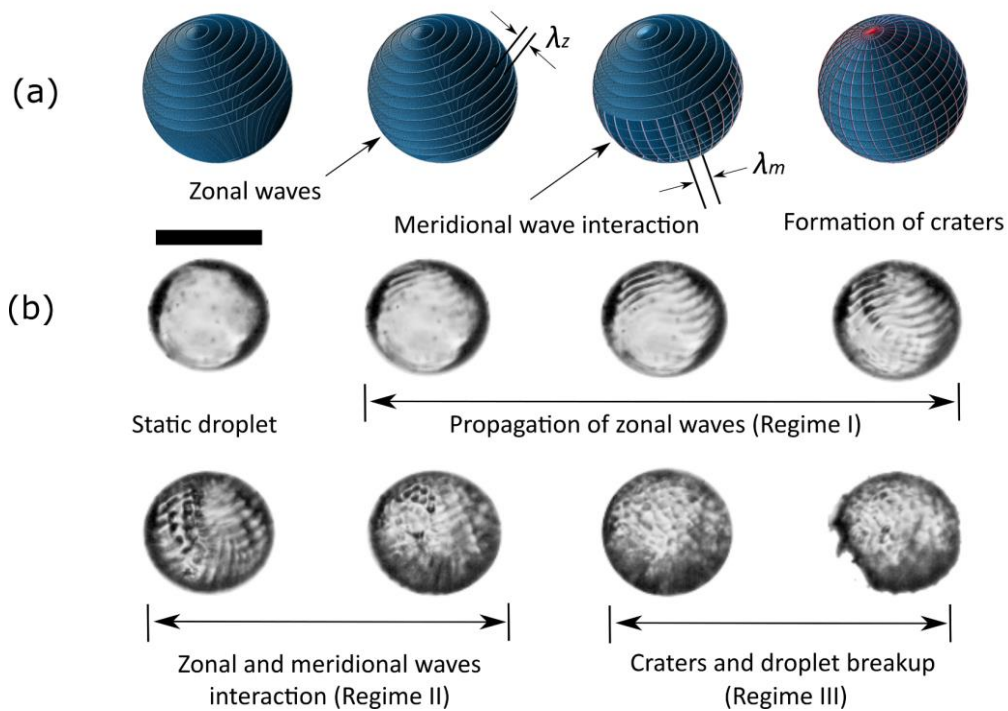

FIG. 8. (a) Schematic of the propagation of capillary waves on the droplet surface. (b) Experimental observation of zonal wave propagation, the interaction of zonal and meridional

waves, and crater formation on the surface of the droplet. The frames are separated by 50 µs. The scale bar indicates 500 µm.

A typical complete sequence of capillary wave propagation is illustrated in Fig. 8(b) with interframe time of 50 µs. The figure shows the final stages of bubble growth, depicting a smooth surface at $\Delta t = 0$ µs. The propagation of the zonal waves (Regime I), its interaction with meridional waves (Regime II), and the formation of distinct craters (Regime III) on the droplet surface can be identified in the figure. Soon after the formation of craters, the unstable oscillations cause the wave crests to tear off from the droplet surface, ultimately leading to the rupture of the bubble and hence the droplet. Similar patterns on the droplet surface comprising of two perpendicular standing waves are usually found at high frequencies (Faraday 1831; Muller et al. 1998; Liu et al. 2019). By neglecting the dissipative effects in the two-dimensional case, the relationship between the wave frequency and wavelength can be expressed by the dispersion relation as (Faber 1995),

$$f_0 = \sqrt{\frac{2\pi\sigma_L}{\lambda_0^3 \rho_L}} \qquad (2)$$

where $\sigma_L$ and $\rho_L$ are the surface tension and the density of the liquid, and $\lambda_0$ is the wavelength of the standing waves. For the experimentally observed wavelength ($45 \pm 5$ µm), the oscillation frequency of the waves is ~ 50 kHz, which is subharmonic to the excitation frequency (100 kHz) of the levitator. This subharmonic response is consistent with the classical experimental observations related to the Faraday instability caused by external vibrations (Faraday 1831).

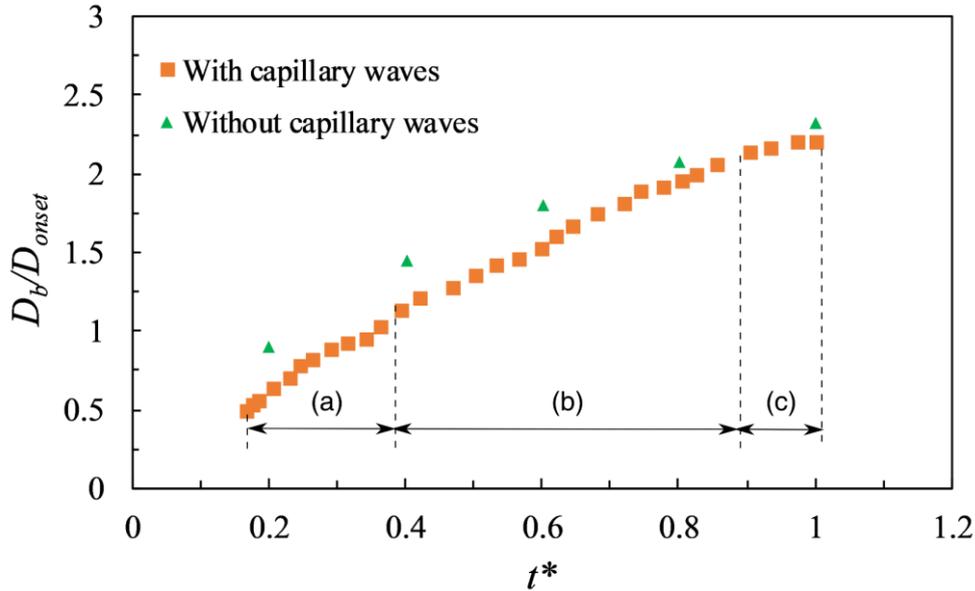

FIG. 9. The temporal evolution of bubble diameter ($D_b$) normalized with the droplet diameter at the onset of the nucleation ($D_{onset}$) for droplets accompanied with and without the capillary waves. The bubble growth consists of three regimes indicating (a) bubble growth without stretching, (b) bubble growth with stretching, and (c) bubble growth with wave propagation.

Also, an excellent correlation is obtained between the capillary wavelength (45 ± 5 µm) and the initial mist of droplets (~ 30 µm) produced by the tearing of the wave crests. Figure 9 represents three regimes in the temporal evolution of vapor bubbles with and without capillary waves. Regime (a) corresponds to the growth of the vapor bubble inside the droplet without any stretching of the liquid layer. Following smooth bubble growth, the vapor bubble stretches the liquid layer at a similar rate (Regime (b)), which is followed by bubble growth with wave propagation (Regime (c)).

### 3.2.2 Bubble breakup

The entire bubble breakup process is divided into three regimes, viz., (1) bubble rupture and cavity formation, (2) stretching (with end pinching), and (3) retraction (with symmetrical pinch-off). It has been well established that when the bubble size is sufficiently large, its collapse leads to the creation of a ligament (Rao et al. 2017a, 2018). During the capillary wave propagation (as discussed in the sec. 3.2.1), the net force acting at the vapor-liquid interface surpasses the restoring surface tension force, resulting in the collapse of the bubble. This bubble collapse leads to the creation of a cavity inside the droplet (see Regime I in Fig. 10), where $\Delta t$ = 0 ms signifies the onset of bubble collapse. The cavity formation is followed by the ejection of a long ligament which stretches with time. The stretching ligament undergoes fragmentation through asymmetric end-pinching mechanism generating small-sized secondary droplets (~ 20 − 50 µm). Here, the ligament stretches from both the open ends into a nearly uniformly shaped cylindrical structure while ejecting fine secondary droplets from only one end (Regime II). Subsequently, the ligaments undergo axisymmetric breakup through capillary wave instabilities producing large-sized secondary droplets (~ 250 µm). This axisymmetric breakup occurs during the retraction of the ligament (Regime III). Once the stretching is slowed down, multiple necking begins to take place. This necking of the thick ligament continuously evolves, while the ends of the ligament form bulbous ends, which gradually detach from the ligament (at $\Delta t$ = 1.45 ms in Fig. 10). Consequently, two more bulbs are formed in the residual ligament; however, they retract towards the ligament center before undergoing pinch-off into secondary droplets (1.75 ms). The formation of multiple necks on the ligament can be attributed to the slightly viscoelastic nature of the emulsion. Moreover, this multi-necking behavior has not been reported in the literature concerning multi-component miscible droplets (Rao et al. 2017c, 2018). We quantified the evolution of ligament aspect ratio for droplets undergoing breakup following capillary wave propagation (NOD ~ 1) and without wave propagation (NOD ~ 0.2).

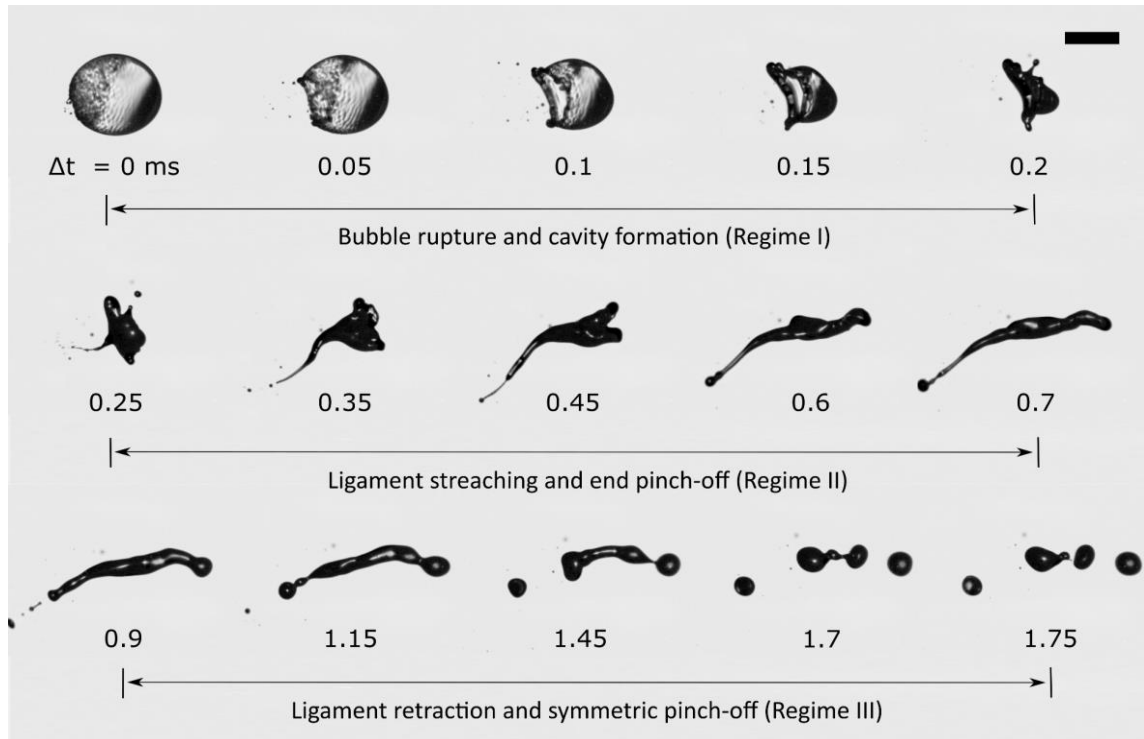

FIG. 10. The rupture of the bubble resulting in the development of long ligament, which consequently undergoes breakup into multiple secondary droplets. The scale bar indicates 500 µm.

Initially, the aspect ratio of the ligament, i.e., the ligament length normalized with maximum ligament diameter ($AR_{lig} = l_{lig}/D_{lig,\ max}$) increases as a function of time, $t/t_b$ (instantaneous time normalized with ligament breakup time) irrespective of whether the breakup is followed by wave propagation or not (Regime II) (see Fig. 11). The ligament reaches maximum extension at $AR_{lig} \sim 9$ before it experiences retraction and consequent pinch-off (Regime III). The extension of the ligament (for $D_{lig} \sim 180$ µm) is directly associated with the liquid surface tension, which tries to constrain the deformation at the capillary timescale, $\tau_c = \sqrt{\rho_L R_{lig}^3 / \sigma_L} = 0.3$ ms, where $\rho_L$ and $\sigma_L$ are the density and surface tension of the liquid respectively. Although the order of the capillary time matches with the experimental ligament breakup timescale (~ 0.85 ms), it is smaller by a factor of 3. The deviation in the capillary time is possibly due to the presence of a surfactant, which plays a role in delaying the breakup of the ligament (Dechelette et al. 2011).

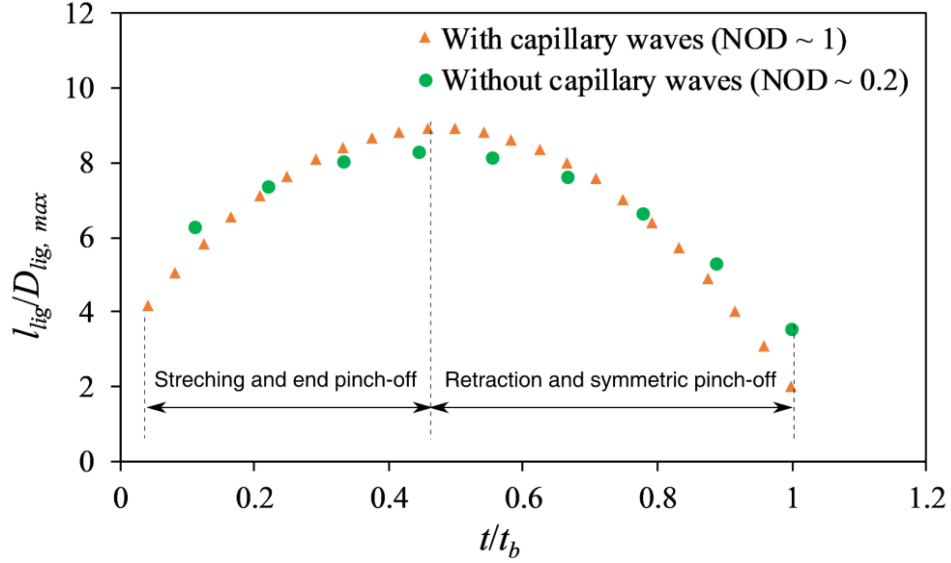

FIG. 11. Evolution of ligament aspect ratio ($l_{lig}/D_{lig,\,max}$) following the bubble growth and breakup accompanied by wave propagation and without wave propagation.

The bubble rupture occurs because of the imbalance in the net force ($F_{net}$) on the vapor-liquid interface and the restoring surface tension ($F_{st}$) of the vapor bubble (with surface area, $A_{bubble}$), i.e., $F_{net} > F_{st}$ (Pathak and Basu 2016b). The net force is given as,

$$F_{net} = (P_{int} - P_{amb})A_{bubble} \qquad (3)$$

Here $P_{int}$ is the vapor bubble pressure and $P_{amb}$ is the ambient pressure. As discussed before, the rupture of the bubble results in the creation of a cavity on the droplet surface. Afterward, this unstable cavity rapidly re-equilibrates due to capillary forces and leads to the growth of the ligament (with radius $R_{lig}$ and length $l_{lig}$) that stretches longitudinally due to the residual energy ($E_{residual}$). The extending ligament further undergoes breakup into several secondary droplets, which may further evaporate and possibly experience secondary breakup. The residual energy required for the ligament growth is given as

$$E_{residual} = \left[ P_{int} - P_{amb} - \left( \frac{F_{st}}{A_{bubble}} \right) \right] \Delta V_{bubble} \qquad (4)$$

The energy balance associated with ligament extension can be described as

$$\frac{1}{2}m_{lig}v_{lig}^2 = E_{residual}, \qquad (5)$$

Here, $m_{lig}\left(=\frac{\pi}{4}D_{lig}^2 l_{lig}\rho_L\right)$ is the mass of the ligament and $v_{lig}\left(=\frac{dl_{lig}}{dt_{lig}}\right)$ is the velocity of the stretching ligament. Therefore, the length scale of the ligament ($L_{lig}$) can be approximated using equation (5) as

$$l_{lig} = \left(\frac{8E_{residual}t_{lig}^2}{\pi D_{lig}^2 \rho_L}\right)^{1/3}, \qquad (6)$$

For the breakup event shown in Fig. 10, the estimated ligament length ($L_{lig}$) at the onset of pinch-off is ~ 2 mm, which agrees very well with the experimental ligament length (~ 1.5 mm). The aspect ratio ($l_{lig}/D_{lig}$) of the ligaments is observed to be higher than $\pi$ (~ 3.14) in all the ligament breakup scenarios, indicating that the ligaments undergo a Rayleigh-Plateau type of breakup.

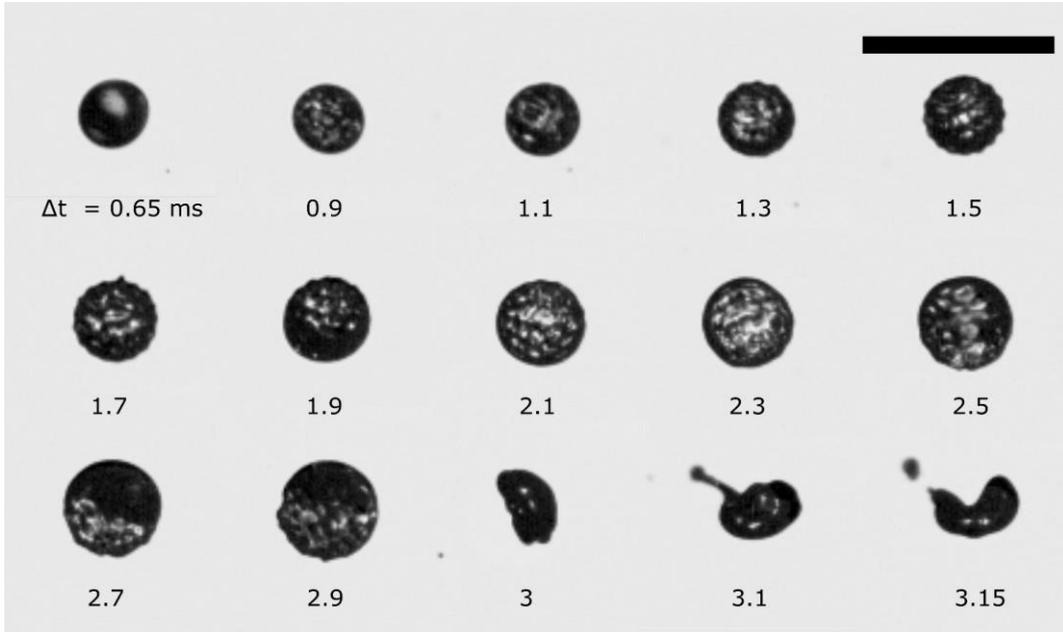

FIG. 12. The growth of the vapor bubble inside the secondary droplet and its subsequent breakup. The scale bar indicates 500 µm.

After the first bubble breakup event, an appropriate proportion of water exists even in the secondary droplets, which is adequate for further bubble growth to take place inside the droplet.

Figure 12 shows the growth of vapor bubble inside the secondary droplet (corresponding to Fig. 10), where $\Delta t = 0$ ms represents the onset of bubble nucleation in the secondary droplet. As seen in the figure, the vapor bubble does not grow significantly larger than the diameter of secondary droplet at the onset of nucleation. Accordingly, the liquid layer surrounding the bubble is not sufficiently thin for the wave propagation to occur due to the domination of both the liquid surface tension and the acoustic pressure over the internal pressure of the bubble. Therefore, the wave pattern on the surface of the droplet (0.9 ms) is triggered by only acoustic levitation without prior capillary wave propagation. The secondary or tertiary breakup events do not possess as much strength as the primary bubble breakup event (3.1 ms), resulting in relatively inefficient atomization compared to the primary breakup event. The cascade of breakup events occurs irrespective of whether the breakup is instability driven or not. For example, Fig. 13 shows the growth of the vapor bubble and its breakup in secondary droplet without the occurrence of Faraday instability on the surface of parent droplet. These multiple breakup events continue until the proportion of water in the secondary droplets is low enough for vapor bubble growth. These multiple breakup events have also been frequently reported during the combustion of multi-component miscible droplets (Rao et al. 2017b, 2018).

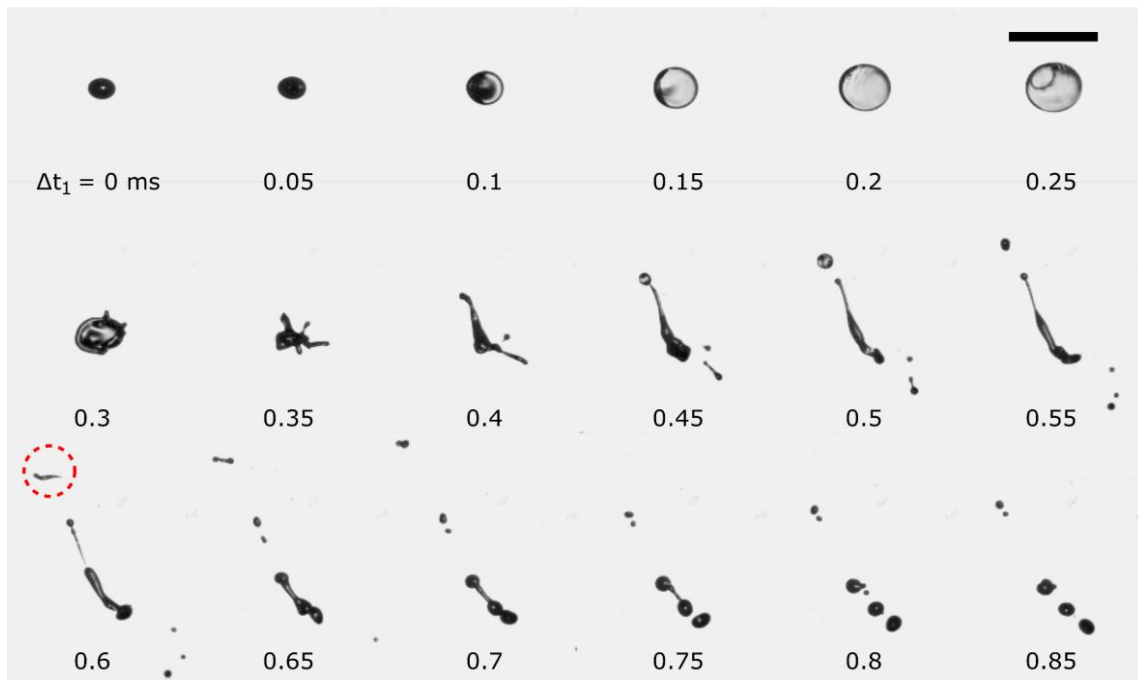

FIG. 13. Vapor bubble growth and breakup without wave propagation on the droplet surface. The red dotted circle indicates the secondary breakup event. The scale bar indicates 500 µm.

It is important to note that a bubble breakup event does not always entirely fragment the parent droplet into multiple secondary droplets. When the strength of bubble breakup is not intense, the parent droplet may experience multiple bubble growth and collapse cycles. These cycles help in depleting a significant proportion of water present in the droplet while the surfactant concentration remains nearly the same. This leads to the formation of a polymer-like structure on the droplet surface, as seen in Fig. 14, where $\Delta t = 0$ ms represents the onset of the formation of dendrite-like structures. Initially, the structures are faintly visible ($\Delta t = 0.5$ ms) and grow gradually, forming complex structures resembling dendrites (1.2 ms). When the structures completely occupy the interface, the bubble collapses (1.45 ms) and grows again. This time, capillary waves begin to propagate on the droplet surface; however, the waves do not cover the entire droplet surface due to the presence of a thick liquid layer surrounding the bubble (1.6 ms). Finally, the bubble and hence the droplet collapses and leads to the ejection of a thick ligament (2.3 ms), which subsequently undergoes pinch-off (2.55 ms).

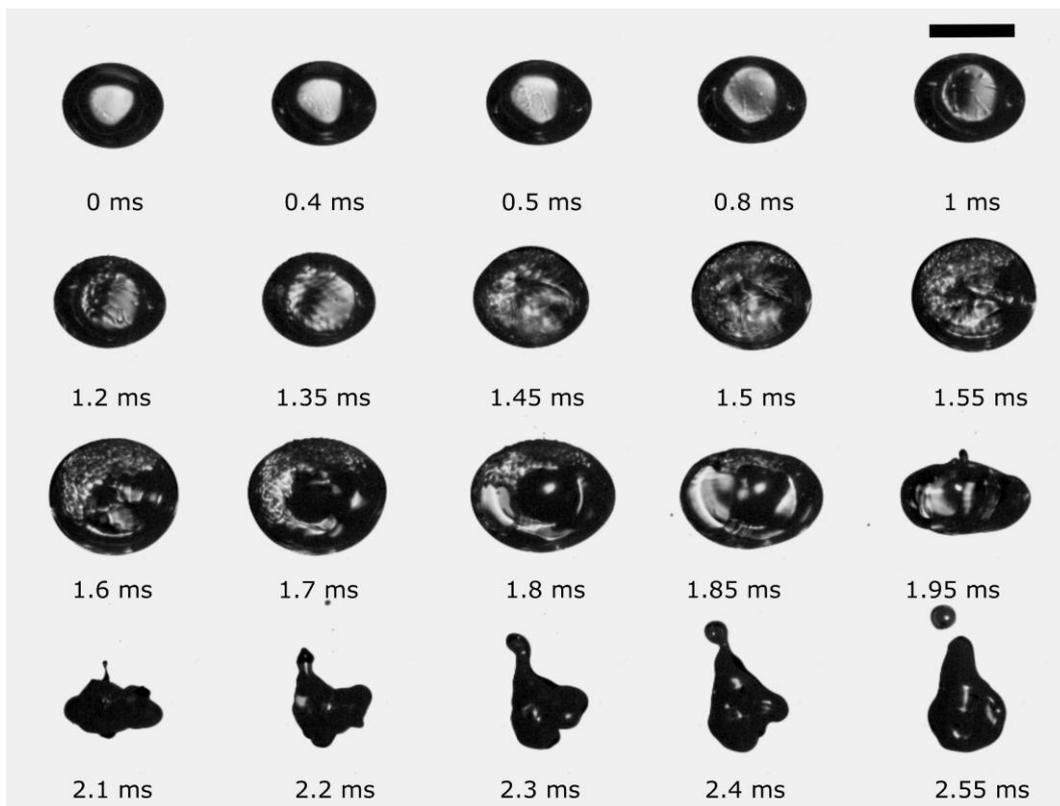

FIG. 14. Formation of a polymer-like structure on the surface of the droplet. The scale bar indicates 500 µm.

### 3.3 Sheet formation and breakup

Sheet formation predominantly occurs in emulsions of relatively high volatility difference (water/dodecane emulsion) than that required for bubble growth (water/decane emulsion). In contrast to the breakup of large bubbles with $\alpha > 1$ (sec. 3.2), sheet formation occurs due to the breakup of an indiscernible small sized bubble ($\alpha < 1$). Moreover, it is highly probable that the diameter of the droplet at the onset of the breakup is relatively smaller (~ 0.2–0.3 mm) compared to the droplet breakup via bubble growth due to the evaporation of droplet before the occurrence of breakup. Prior to the formation of the liquid sheet, fragments of droplets are ejected upon the breakup of parent droplet (see $\Delta t = 0.05$ ms in Fig. 15). Here, $\Delta t = 0$ ms represents the onset of droplet breakup. As the sheet propels in the opposite direction due to the reactionary thrust, it develops a circular rim. Owing to the instability on the sheet rim, ligaments are formed, which subsequently undergo pinch-off, and the secondary droplets are ejected from the circular rim of the sheet. Both the rim, as well as the ligaments attached to the sheet, are visible in Fig. 15 at 0.05 ms, which becomes more pronounced with time (0.1 ms). The sheet soon collapses under the influence of surface tension (0.15 ms) and transforms into a long ligament (0.3 ms). The ligament then becomes Rayleigh-Plateau unstable and undergoes pinch-off into several secondary droplets.

It is observed that the lifetime of the sheet increases with the increase in the diameter of the droplet at the onset of the breakup. For example, when the droplet size at the onset of the breakup is relatively large (~ 0.3 mm) during the evaporation process (although with marginal reduction in size), a stable sheet is formed that remains intact for an extended period (~ 0.4 ms). As the sheet expands, the initially expelled fragments impinge on the hemispherical sheet (see Fig. 16). The fragments which are about to impact the sheet are highlighted by the dotted rectangle at $\Delta t = 0.15$ ms. Due to the impact of the fragments on the thin sheet, capillary waves travel on the sheet (highlighted by dotted circle at 0.2 ms and 0.25 ms). The expanding sheet reaches its maximum extension at approximately 0.2 ms and, thereafter, collapses under the influence of surface tension.

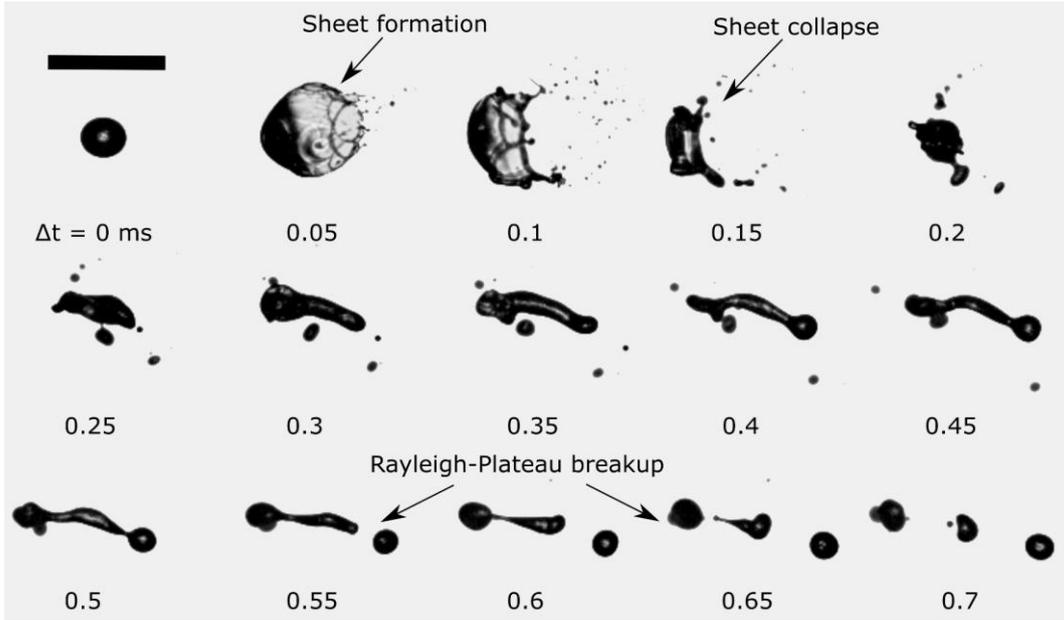

FIG. 15. Crown formation and the breakup of an emulsion droplet ($D_{onset} \sim 0.2$ mm) followed by the ligament-mediated breakup. The scale bar indicates 500 µm.

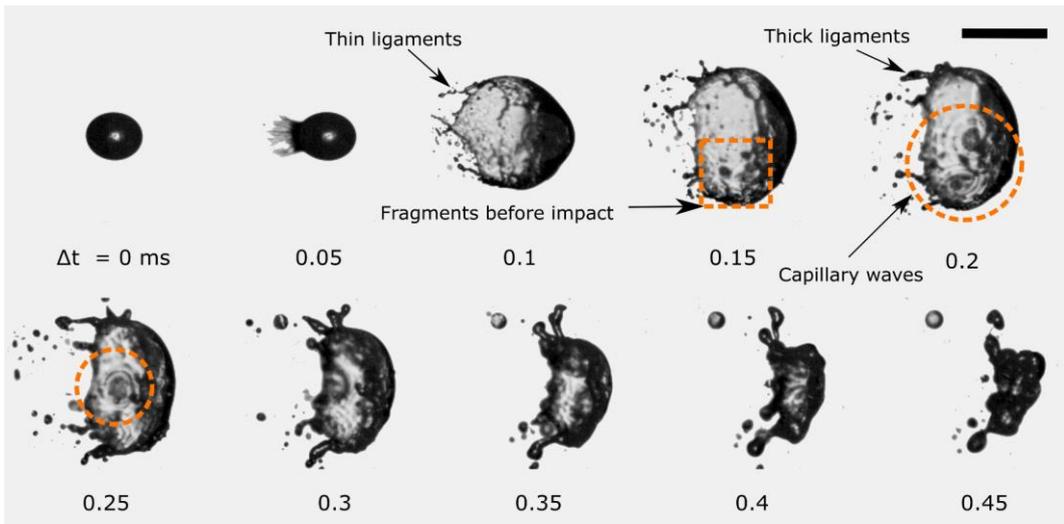

FIG. 16. Crown formation and the breakup of an emulsion droplet ($D_{onset} \sim 0.3$ mm) followed by capillary wave propagation on the sheet surface. The scale bar indicates 500 µm.

By equilibrating the kinetic energy of the sheet with the surface energy, and considering an insignificant mass loss from the droplet at the onset of breakup, the capillary time can be expressed as (Gonzalez Avila and Ohl 2016)

$$\tau_c = \sqrt{\frac{\rho_L R_{onset}^3}{6\sigma_L}} \quad (7)$$

For the droplet depicted in Fig. 16, we calculate $\tau_c \sim 0.21$ ms, which agrees well with the experimentally observed sheet lifetime (0.25 ms).

Soon after the onset of droplet fragmentation, the sheet is exposed to significant radial acceleration. The acceleration of the sheet ($a_s$) estimated from Fig. 16 is of the order of $10^5$ m/s$^2$. Therefore, the expanding sheet is subject to Rayleigh–Taylor (RT) instability (Gonzalez Avila and Ohl 2016), where the instability growth can be estimated from Villermaux and Clanet (2002)

$$\Delta t_{RT} = \left( \frac{\sigma_L}{\rho_L a_s^3} \right)^{1/4}, \qquad (8)$$

For a DDW30 droplet shown in Fig. 16, the growth rate is $\Delta t_{RT} = 20$ μs. Although the interframe interval of 50 μs did not allow us to provide a lower limit of the onset of the instability, the order of $\Delta t_{RT}$ is comparable. As the crown sheet decelerates, a gradual accumulation of the liquid takes place at the sheet edge. As seen in Fig. 15 (0.05 ms) and Fig. 16 (0.1 ms), this continuous accumulation of liquid results in the creation of a rim. In both the sheets, ligaments are attached to the rim owing to the Rayleigh–Taylor instabilities. The evolution of the rim and the ligaments is visible in Fig. 16 where initially (at 0.1 ms), both the rim and the ligament attached to it are thin (highlighted by arrow). This ligament develops into a pronounced rim at 0.35 ms. These ligaments become Rayleigh–Plateau unstable and eject droplets from the edge of the crown sheet. These droplets are much bigger than the fine mist, which is expelled soon after the onset of fragmentation. The side view of the expansion and breakup of the crown sheet is shown in Fig. 17 (low probable case) for a droplet with even larger $D_{onset}$ (~ 0.4 μm), where the sheet expands and reverses its direction to the reactionary thrust (Multimedia view). As expected, the sheet expansion is prolonged to approximately 0.6 ms. A similar sheet expansion has also been reported for a droplet impacted by a laser pulse (Klein et al. 2015).

Besides the formation of stable sheets, unstable sheet and its breakup have also been observed, which is characterized by the formation of patches and holes across the sheet (Rao et al. 2019). However, the probability of its occurrence is very low. Similarly, the droplet also undergoes bag-type breakup; however, its behavior is stochastic compared to other fragmentation modes.

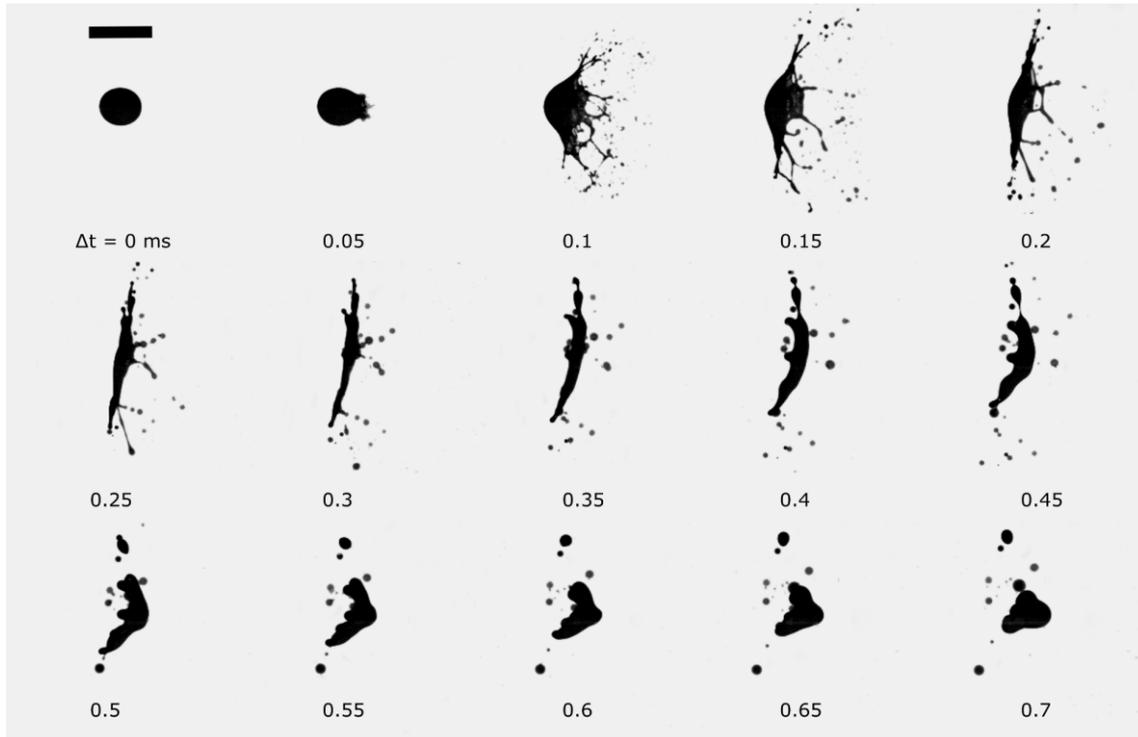

FIG. 17. Expansion of crown sheet and its reversal due to the breakup-induced reactionary thrust. The scale bar indicates 500 µm.

### 3.4 Catastrophic breakup

Catastrophic mode of breakup leads to an abrupt transition of a smoothly evaporating droplet into prompt fragmentation. It can be contemplated that in this breakup mode, the size of the bubble is significantly smaller compared to the droplet diameter at the onset of breakup ($\alpha \ll 1$). This mode of breakup primarily occurs for emulsions having components with significantly vast volatility difference (i.e., water/tetradecane emulsions). Since the lower volatile component drives the temperature of the droplet surface, the time required to superheat the higher volatile component is substantial. This delay in superheating leads to significant evaporation and regression of droplet diameter (see Fig. 6). Therefore, the droplet size at the onset of breakup in these mixtures is substantially smaller than the previously discussed emulsified droplets (Sec. III B and III C). However, the delay in superheating, in turn, leads to a very high degree of superheating of the water. The breakup of the high-pressure vapor bubble disintegrates the parent droplet and ejects the secondary droplets in all directions at a high velocity (~ 10 m/s). In this breakup mode, the droplet explodes into vapor and fine secondary droplets with an audible "popping" sound. The tremendous strength of catastrophic breakup leads to the complete fragmentation of droplet within 0.2 ms. Figure 18(a) shows a sequence of a catastrophic breakup event (unstable prompt fragmentation), which appears to be the transition of breakup from sheet to catastrophic breakup (with secondary droplet size ranging ~ 20 − 50 µm). Whereas, Fig. 18(b) shows a complete catastrophic breakup where the droplet disintegrates abruptly, leading to the generation of uniformly sized secondary droplets (~ 20 µm). The intense fragmentation results in the formation of a vapor cloud, as seen in Fig. 18(b) at 0.2 ms (Multimedia view).

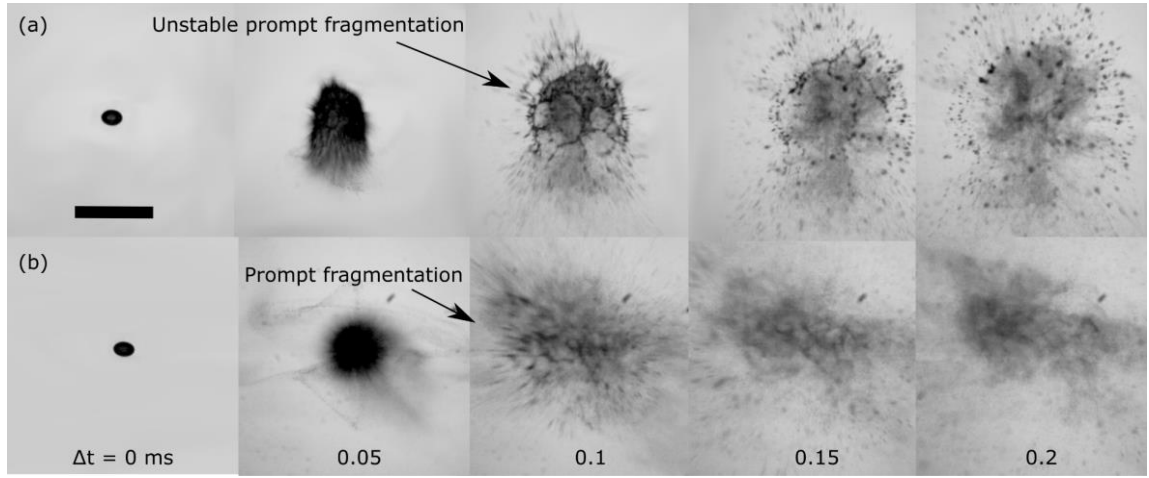

FIG. 18. (a) The transition from sheet formation to catastrophic breakup. (b) Catastrophic breakup. The scale bar indicates 500 µm.

A summary of the different breakup configurations is shown in Fig. 19. The three breakup scenarios, namely, breakup through bubble growth, sheet breakup, and catastrophic breakup, are plotted with distinct symbols, and they are depicted with a characteristic experimental image. The ratio of the force generated by the bubble ($F = (P_{int} - P_{amb})A_{bubble} = M_{res}a_{res}$) to the surface tension force of the liquid ($F_{st} = 2\pi R_{onset}\sigma_L$) provides an overall criterion for the occurrence of different modes of breakup regimes. Note that $F$ is different from $F_{net}$ mentioned in section 3.2. Here $A_{bubble}$ is the area of the vapor bubble, which cannot be accurately discerned in all the breakup scenarios. $M_{res}$ and $a_{res}$ are the mass and average acceleration of the residual droplet or fragments (see section 2), and $\sigma_L$ is the surface tension of liquid at wet bulb temperature.

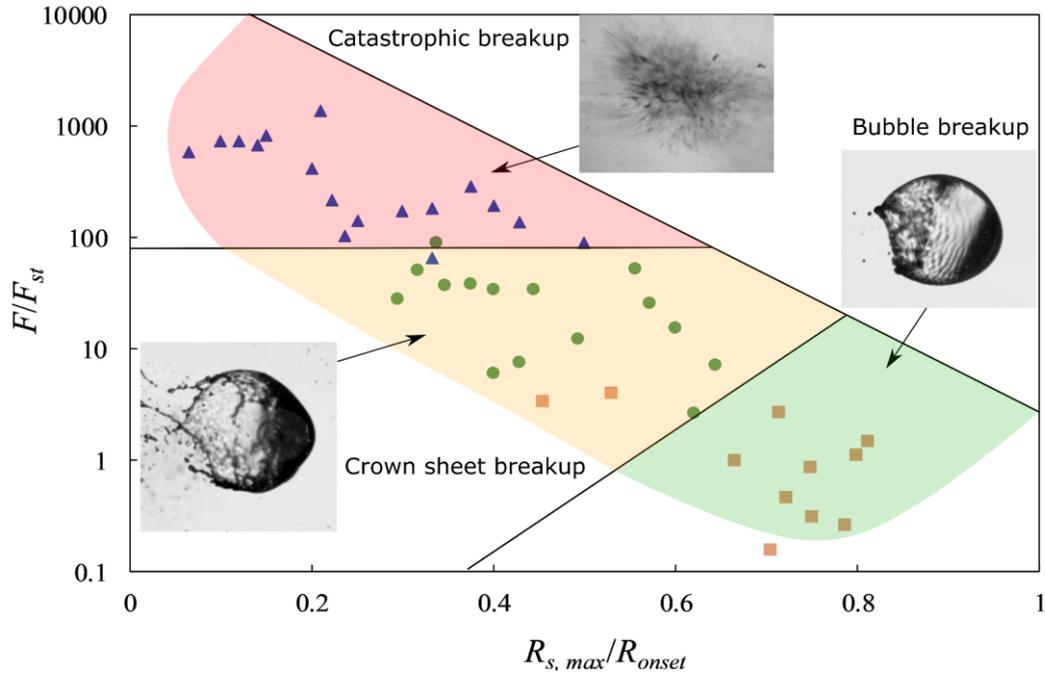

FIG. 19. Overview of the breakup configurations of water-in-fuel emulsions studied in the present work.

When $F/F_{st}$ is small (up to ~ 10), production of secondary droplets occurs due to bubble breakup, which is followed by ligament-mediated breakup through Rayleigh-Plateau instability (most probable in water/decane emulsions). The secondary droplets formed through this mode of breakup are large-sized ($R_{s,\,max}/R_{onset}$ ~ 0.6 − 0.8) compared to other modes of breakup, where $R_{s,\,max}$ is the maximum radius of the secondary droplet. At relatively higher $F/F_{st}$ (up to ~ 100), the droplet breaks into slightly smaller secondary droplets ($R_{s,\,max}/R_{onset}$ ~ 0.3 − 0.6) through the breakup of the crown sheet (most probable in water/dodecane emulsions). Finally, when the droplets undergo catastrophic breakup, a fine mist of secondary droplets is produced ($R_{s,\,max}/R_{onset}$ ~ 0.1 − 0.4), resulting in the most efficient atomization among all the emulsions studied in the present work (most probable in water/tetradecane emulsions). It is important to note that the position of the indiscernible vapor bubble inside the parent droplet can significantly influence the regime, which may explain the scatter seen in Fig. 19. Nonetheless, a simple regime map is shown, which should be verified with further experimentations and numerical simulations.

# 4 Conclusions

We have analyzed different breakup scenarios of an evaporating emulsion droplet, and the conclusions derived from the present work are as follows:

1. Three broad breakup mechanisms are revealed in the present work, (1) Breakup through bubble growth, (2) sheet breakup, and (3) catastrophic breakup. The occurrence of these breakup modes is found to be mainly dependent on the onset of vapor bubble nucleation, which in turn was found to dictate the strength of breakup. It is also shown that the types of breakup differ from each other not only in their external appearance but also in the spectra of secondary droplets formed upon breakup.

2. For the droplets undergoing breakup through bubble growth, intricate patterns of wave propagation are observed on the droplet surface for the first time (Most probable in water/decane emulsions). These patterns are, in turn, excited by the vertical vibration due to the presence of acoustic field induced resonance. The short time scale instability on the droplet surface is found to be the manifestation of Faraday instability.

3. The deformation and breakup of droplets leading to sheet formation predominantly occur in emulsions consisting of components with relatively higher volatility difference (water/dodecane emulsions). Sheet formation occurs due to the breakup of an indiscernible small sized bubble ($< 10\,\mu m$). These stable sheets do not undergo efficient atomization upon their collapse.

4. Catastrophic breakup primarily occurs for emulsions with significantly vast volatility difference (water/tetradecane emulsions). The droplet size at the onset of breakup in these mixtures is substantially smaller compared to other emulsions.

5. Among all the modes of breakup, catastrophic breakup was observed to produce the smallest diameter secondary droplets. The fine secondary droplets ($\sim 10\,\mu m$) produced from a catastrophic breakup could be beneficial in spray combustors using emulsified droplets.

**Conflict of interest**

The authors declare that there is no conflict of interest.